\documentclass[12pt, a4paper, twoside]{amsart}

\usepackage[english]{babel}

  \usepackage{amsmath, amsthm, amssymb, amsfonts}
  \usepackage[colorlinks,backref]{hyperref}
  \usepackage{graphicx}
  \usepackage{color}
  \usepackage{array}
  \usepackage{float}
    \floatstyle{plaintop}
    \restylefloat{table}
    \restylefloat{figure}

  \newcommand{\zapomen}[1]{}
  \newcommand{\TODO}   [1]{\TC{TODO \expandafter\MakeUppercase{#1} TODO}}
  \DeclareMathOperator*{\E}{E}
  \DeclareMathOperator*{\V}{var}
  \DeclareMathOperator*{\Cov}{cov}

  \definecolor {Red}           {cmyk} {0,0.89,0.94,0.28}
  
  \newcommand{\TC}     [1]{\textcolor{Red}     {\textbf{#1}}}

  \newcount \hour
  \newcount \minute
  \hour   = \time \divide \hour by 60
  \minute = \time
  \loop  \ifnum \minute  > 59 \advance  \minute   by -60 \repeat
  \def\wt{\number\hour:\number\minute}
\begin{document}

\title{New randomized response technique for estimating the population
       total of a quantitative variable}

  \author{Jarom\'\i{}r Antoch$^{1,3}$}
  \author{Francesco Mola$^2$}
  \author{Ond\v{r}ej Voz\'{a}r$^{3,4}$}

\address{${}^{1}$\,Charles University, Faculty of Mathematics and
         Physics,
         Sokolovsk\'{a} 83, CZ\,--\,186\,75 Praha~8\,--\,Karl\'\i{}n,
         Czech Republic; antoch@karlin.mff.cuni.cz}

\address{$^2$\,Universit\`a di Cagliari, Facolt\'a di Economia,
         viale S.~Ignazio da Laconi 17, I\,--\,09123 Cagliari, Italy;
         mola@unica.it}

\address{$^3$\,Prague University of Economics and Business, Faculty of
         Informatics and Statistics,
         W.~Churchill Sq.~4, CZ\,--\,130\,67  Praha~3}

\address{$^4$\,Czech Statistical Office,
         Na Pades\'{a}t\'{e}m 3268/81, CZ\,--\,100\,82 Praha~10,
         Czech republic; vozo01@vse.cz}


\begin{abstract}
  In this paper, a new randomized response technique aimed at protecting
  respondents' privacy is proposed. It is designed for  estimating the
  population total, or the population mean, of a quantitative
  characteristic. It provides a~high degree of protection to the
  interviewed individuals, hence it may be favorably perceived by them and
  increase their willingness to cooperate. Instead of revealing the true
  value of the characteristic under investigation, a respondent only states
  whether the value is greater (or smaller) than a~number which is selected
  by him/her at random, and is unknown to the interviewer. For each
  respondent this number, a sort of individual threshold, is generated as a
  pseudorandom number from the uniform distribution. Further, two
  modifications of the proposed technique are presented. The first
  modification assumes that the interviewer also knows the generated random
  number. The second modification deals with the issue that, for certain
  variables, such as income, it may be embarrassing for the respondents to
  report either high or low values. Thus, depending on the value of the
  pseudorandom lower bound, the respondent is asked different questions to
  avoid being embarrassed. The suggested approach is applied in detail to
  the simple random sampling without replacement, but it can  also be
  applied to many currently used sampling schemes, including cluster
  sampling, two-stage sampling, etc. Results of simulations illustrate the
  behavior of the proposed procedure.

\end{abstract}

\maketitle
\markboth{RRT for estimating population total}
         {Antoch, Mola, Voz\'ar : arXiv : \today,\ \wt}

\section{Introduction}\label{introduction}

A steady decline in response rates  has been reported for many surveys in
most countries around the world; see, e.g., Stoop (2005), Steeh et al.
(2001) or Synodinos and Yamada (2000). This decline is observed regardless
of the mode of the survey, e.g., face-to-face survey, paper/electronic
questionnaire, Internet survey or telephone interviewing. Furthermore, this
trend has continued despite additional procedures aimed at reducing refusal
and increasing contact rates $\big($Brick $(2013)\big)$. For some time we
have observed that people are getting more and more suspicious with respect
to any kind of sampling surveys,  a priori assuming that the other side
cheats (or can cheat). It is especially due to the overall spread of the
Internet, where we communicate with anonymous computer robots, leaving us
no chance to check their trustworthiness.

The growing concern about ``invasion of privacy'' thus also represents an
important challenge for  statisticians. Quite naturally, a respondent may
be hesitant or even evasive in providing any information which may indicate
a deviation from a social or legal norm, and/or which he/she feels might be
used against him/her some time later. Therefore, if we ask sensitive or
pertinent questions in a survey, conscious reporting of false values would
often occur $\big($S\"arndal et al. (1992), pp~$547\big)$. Unfortunately,
standard techniques such as reweighting or model-based imputation cannot
usually be applied; for a thorough discussion, see S\"{a}rndal et al.
(1992) or S\"arndal and Lundstr\"om (2005). On the other hand, this issue
can, at least partially, be resolved using randomized response techniques
(RRT).

For all of the  reasons mentioned above, different RRTs have been developed
with the goal to obtain unbiased estimates and to reduce the non-response
rate. These techniques started with a seminal paper by Warner (1965), who
aimed at estimating the proportion of people in a given population with
sensitive characteristics, such as substance abuse, unacceptable behavior,
criminal past, controversial opinions, etc. Eriksson  (1973) and Chaudhuri
(1987) modified Warner's method to estimate the population total of a
quantitative variable. However, in our opinion based on personal practical
experience,  these standard RRTs aimed at estimating the population total
are rather complicated and demanding on both respondents and survey
statisticians for various real life applications, see also the discussion
in Chaudhuri (2017). They require ``non-trivial arithmetic operations''
from respondents within Chaudhuri's approach, while the survey statistician
must expend a lot of effort connected with the design of a suitable deck of
cards, or other randomization mechanisms to be used for masking the
sensitive variables (such as income, personal wealth) and, at the same
time,  providing accurate enough estimates.

In this paper, we propose a method which is simpler in comparison with
those proposed previously and is practically applicable. The respondent is
only asked whether the value of a sensitive variable attains at least a
certain random lower bound. This technique, and its modifications, are
developed in detail, applied to the simple random sampling without
replacement, and illustrated using simulations.

The main advantages of the suggested method include the ease of
implementation, simpler use by the respondent, and practically acceptable
preciseness. Moreover, respondents' privacy is well protected because they
never report the true value of the sensitive variable. Unlike in
Chaudhuri's or Eriksson's approach, there is no issue with the cards
design. From a certain point of view, a small disadvantage may be a lower
degree of confidence in anonymity, due to the extrinsic device/technique
used for generating random numbers.

This paper is organized as follows. In  sec.~\ref{prehled}, selected
randomized response techniques for estimation of the population total, or
population mean, are concisely described. In sec.~\ref{newProcedure}, a new
randomized response technique and its two modifications are proposed, and
their properties studied. Sec.~\ref{simulations} illustrates the suggested
ideas with the aid of a simulation study. Finally, sec.~\ref{conclusions}
provides the main conclusions of the paper.

\section{Selected randomized response techniques for estimating the
         population total and their properties}\label{prehled}

Let us consider a finite population $U=\{1,\dots,N\}$ of $N$~identifiable
units, where each unit can unambiguously be identified by its label.
Let~$Y$ be a sensitive quantitative variable; the goal of the survey is to
estimate the population total $t_{Y}=\sum_{i \in U} Y_{i}$ or,
alternatively, the population mean $\overline t_{Y}=t_Y/N$, of the surveyed
variable. To that end, we use a random sample~$s$ selected with
probability~$p(s)$, described by a sampling plan with a fixed sample
size~$n$. Let us denote by $\pi_i$ the probability of inclusion of the
$i^{th}$ element in the sample, i.e., $\pi_{i}=\sum_{s \ni i} p(s)$, and by
$\xi_{i}$  the indicator of inclusion of the $i^{th}$ element in the
sample~$s$, i.e., $\xi_{i}=1$ if $s \ni i$ and $\xi_{i}=0$ otherwise. To
keep the length of the paper acceptable, we do not introduce all notions
from scratch and refer the reader to Till\'e (2006) if needed.

As argued above, in practice  it is often impossible to obtain the values
of the surveyed variable~$Y$ in sufficient quality because of its
sensitivity. Therefore, statisticians try to obtain from each respondent at
least a randomized response~$R$ that is correlated to~$Y$. Randomization of
the responses is carried out independently for each population unit in the
sample.

Note that, in such a case, the survey has two phases. First, a sample~$s$
is selected from~$U$ and then, given~$s$, responses $R_i$ are realized
using the selected RRT. We denote the corresponding probability
distributions by $p(s)$ and $q\big(r\,|s\big)$. In this setting, the
notions of the expected values, unbiasedness and variances are tied to a
twofold averaging process:

\begin{itemize}
  \item Over all possible samples~$s$ that can be drawn using the selected
        sampling plan $p(s)$.

  \item Over all possible response sets~$r$ that can be realized
        given~$s$ under the response distribution $q\big(r\,|s\big)$.

\end{itemize}

\noindent
Below we follow the literature and, where appropriate, denote the
expectation operators with respect to these two distributions by~$\E_p$
and~$\E_q$, respectively.

In a direct survey, the population total~$t_Y$ is usually estimated from
the observed values $Y_{i}$ using a linear estimator $t(s,Y)=\sum_{i \in s}
b_{si} Y_{i}$, where the weights $b_{si}$ follow the unbiasedness
constraint $\sum_{s \ni i} p(s)b_{si}=1,\ i=1,\dots,N$. If $\pi_{i}>0\
\forall i \in U$, then Horvitz-Thompson's estimator

\begin{equation}\label{HTestimatorDirect}
  t_{HT}(s,Y)=\sum_{i\in s} \frac{Y_i}{\pi_i}
\end{equation}

\noindent
is a linear unbiased estimator with the weights $b_{si} = 1/\pi_{i}$, and
$\E_{p}\big(t_{HT}(s,Y)\big)=t_{Y}$, see  Horvitz and Thompson (1952), or
sec. 2.8 in Till\'{e} (2006) for details.

If the survey is conducted by means of RRT, the true values of~$Y_i$ for
the sample~$s$ are unknown and, instead of them, values of random
variables~$R_i$ correlated to~$Y_i$ are collected. The population total is
then usually estimated using a Horvitz-Thompson's type estimator

\begin{equation}\label{HTestimatorUndirect}
  t_{HT}(s,R) =\sum_{i \in s}\frac{R_{i}}{\pi_{i}}.
\end{equation}

Suppose now that we have an estimator (a~formula, or a~computational
procedure) for estimating the population total $t_Y$ or population mean
$\overline t_Y$; we denote it by $\widehat Y_{R}$ and $\widehat{\overline
Y}_{\kern-2pt R}$, respectively. The subscript~$R$ emphasizes that the
estimator is based on the values of~$R_i$ in the sample, i.e., on
randomized responses. Moreover, we assume that the randomized
responses~$R_i$ follow a model for which it holds
$\E\big(R_{i}\big)=Y_{i}$, $\V\big(R_{i}\big)=\phi_{i}\ \forall i\in U$,
and $\Cov \big(R_{i},R_{j}\big)=0\ \forall i\neq j,\ i,j\in U$. Note that
$\phi_{i}$ is a function of $Y_{i}$.

Recall that the estimator $\widehat Y_{R}$ of the population total~$t_Y$
\textit{is conditionally unbiased} if the conditional expectation of
$\widehat Y_{R}$ given the sample~$s$ is equal to the current estimator
$\widehat Y_s$ that would be obtained if no randomization took place, i.e.,
if $\E_q\big(\widehat Y_R\,|\,s\big)= \widehat Y_s$. The subscript~$s$
indicates that the ``usual'' estimator based on the non-randomized sample,
e.g., the Horvitz-Thompson's one, is used, and $\E_q\big(\widehat
Y_{R}\,\big|\,s\big)$ stands for the  conditional expectation of $\widehat
Y_{R}$ given the sample~$s$ with respect to the distribution induced by the
randomization of responses. For the estimator $\widehat{\overline
Y}_{\kern-2pt R}$ of the population mean, we proceed analogously.

If $\widehat Y_{R}$ is conditionally unbiased and $\widehat Y_s$ is
unbiased, then $\widehat Y_{R}$ is unbiased as well, since it holds
$\E{}\big(\widehat Y_{R}\big) = \E{}_p\big(\E{}_q (\widehat
Y_{R}\,\big|\,s)\big)= \E{}_p\big(\widehat Y_s\big)=t_Y$. Analogously
it holds $\E\big(\widehat{\overline Y}_{R}\big) = \overline t_Y$.

By a~standard formula of probability theory, we get the variance
of~$\widehat Y_{R}$ in the form

\begin{align}\label{varianceFormula}
   \V{}\big(\widehat Y_{R}\big)=
 & \E{}_p\Big(\V{_q}\big(\widehat Y_{R}\,\big|\,s\big)\Big)+
   \V{}_p\Big(\E{}_q\big(\widehat Y_{R}\,\big|\,s\big)\Big)\notag\\
 &=\E{}_p\Big(\V{_q}\big(\widehat Y_{R}\,\big|\,s\big)\Big)+
   \V{_p}\big(\widehat Y_s\big).
\end{align}

The second term on the right-hand side of~\eqref{varianceFormula} is,
obviously, the variance of the estimator that would apply if no
randomization of responses was deemed necessary, while the first term
represents the increase of the variance produced by the randomization. In
other words, the two terms on the right-hand side
of~\eqref{varianceFormula} represent, respectively, ``\textit{contribution
by randomized response technique used}'' and ``\textit{contribution by
sampling variation}'' to the total variance of $\widehat Y_{R}$. When
treating $\widehat{\overline Y}_{\kern-2pt R}$, we proceed analogously.

Because the variances of $\widehat Y_s$ are well known for many currently
used sampling procedures, it remains to find the contribution by
randomization and to suggest methods for its estimation.

For the estimator~$t_{HT}(s,R)$ given by~\eqref{HTestimatorUndirect}, we
have

\begin{align}\label{varianceDecomposition}
     \V{}\Big(t_{HT}(s,R)\Big)
  &= \E{}_{p}\Big(\V{}_q\big(t_{HT}(s,R)\,\big|\, s\big)\Big)
    +\V{}_p\Big(\E{}_q\big(t_{HT}(s,R)\,\big|\, s\big)\Big) \notag\\
  &= \E{}_{p}\Big(\sum_{i \in U}\frac{\phi_{i}\xi_{i}}{\pi^{2}_{i}}\Big)
    +\V{_p}\big(t_{HT}(s,Y)\big) \notag\\
  &=\sum_{i \in U}\frac{\phi_{i}}{\pi_{i}}
    +\V\big(t_{HT}\big(s,Y\big)\big).
\end{align}

When any RRT is used instead of direct surveying, the variance of the
population total estimator is always higher. This increase in variability
of $t_{HT}(s,R)$ is described by the first term in
\eqref{varianceDecomposition}, which represents additional variability
caused by using a randomized response~$R$ instead of the directly surveyed
variable~$Y.$

Let us take a look at two RRT proposals that are recommended in the
literature and used in  practice. Note that the subscript~$E$,
$C$~(respectively) emphasizes Eriksson's, Chaudhuri's (respectively)
approach; each of them is concisely revisited below.

Eriksson (1973) proposed a technique in which the respondent randomly draws
a card from a deck. The deck contains $100\,C$\%, $0<C<1$, cards with the
text ``\textit{True value}'', while the remaining cards have values
$x_{1},\dots,x_{T}$ with relative frequencies $q_{1},\dots,q_{T}$,
$\sum_{t=1}^{T} q_{t}=1-C$. The values of cards $x_{1},\dots,x_{T}$ are
chosen to mask the true values of the surveyed variable~$Y$. Each respondent
randomly draws one card from a deck. If a card with the text ``\textit{True
value}''  is selected, then the true value of~$Y$ is reported, otherwise
the value $x_{t}$ shown on the card is given. The respondent then returns
the selected card to the deck, and the interviewer does not know which card
it was. The answer from the $i^{th}$ respondent is thus a random variable

\[
   Z_{i,E} =
  \begin{cases}
    Y_{i}, & \textrm{with probability }   C, \\
    x_{t}, & \textrm{with probability } q_{t},\ t=1,\dots,T.
  \end{cases}
\]

\noindent
The answer $Z_{i,E}$ from the $i^{th}$ respondent is then transformed to
$R_{i,E}=\frac{Z_{i,E}-\sum_{t=1}^{T} q_{t}x_{t}}{C}$. It follows from the
definition of $Z_{i,E}$ that the transformed randomized responses $R_{i,E}$
have the expectation and variance values

\begin{align*}
  \E\big(R_{i,E}\big)&=Y_{i}, \\
  \V\big(R_{i,E}\big)&=\frac{C(1-C)Y^2_{i}+\sum_{t=1}^{T} q_{t}x^2_{t}
 -\big(\sum_{t=1}^{T}q_{t}x_{t}\big)^2-2CY_i\sum_{t=1}^Tx_tq_t}{C^2},
\end{align*}

\noindent
so that the corresponding Horvitz-Thompson's type estimator is unbiased.
Unfortunately, if the value reported by a respondent differs from any of
$x_{t}$, the interviewer can deduce the true value of the sensitive
variable; this fact may decrease the credibility and   the willingness of
some respondents to cooperate.

Later on, Chaudhuri (1987) suggested that two decks of cards should be
used. The first deck contains cards with values $a_{1},\dots,a_{K}$, and
the second deck values $b_{1},\dots,b_{L}$. Both decks of cards should mask
the behavior of the studied variable~$Y$. Moreover, the following
relationships must hold:

\begin{align*}
  \mu_{a} &= \frac1{K}\sum_{k=1}^{K}a_{k} \neq 0,
  &\sigma^{2}_{a}=\frac1{K}\sum_{k=1}^{K}(a_{k}-\mu_{a})^2>0, \\
  \mu_{b} &= \frac1{L}\sum_{l=1}^{L}b_{l} \neq 0,
  &\sigma^{2}_{b}=\frac1{L}\sum_{l=1}^{L}(b_{l}-\mu_{b})^2>0.
\end{align*}

The respondent randomly draws one card from each deck, say $a_k$ and $b_l$,
whereas the interviewer does not know the values on the drawn cards. Then
the respondent returns both cards, and instead of the true value $Y_{i}$
the value of $Z_{i,C}=a_{k}Y_{i}+b_{l}$ is reported. This response is then
transformed to the randomized response $R_{i,C}=\frac{Z_{i,C}-\mu_{b}}
{\mu_{a}}$. It follows from the definition of $Z_{i,C}$ that the randomized
response $R_{i,C}$ has the expectation and variance

\[
  \E\big(R_{i,C}\big)=Y_{i} \quad \textrm{and} \quad
  \V\big(R_{i,C}\big)=Y_{i}^2\frac{\sigma_{a}^2}{\mu_{a}^2}+
  \frac{\sigma_{b}^2}{\mu_{a}^2},
\]

\noindent
so that corresponding Horvitz-Thompson's type estimator is also unbiased.

Both Eriksson's and Chaudhuri's techniques have been further developed and
improved by other researchers, see, e.g., an interesting papers by Arnab
(1995, 1998), Gjestvanga and Singh (2009) or Bose and Dihidar (2018). The
ideas and a representative review of further research are presented in a
monograph by Chaudhuri (2017). Other types of randomization techniques were
suggested in a series of papers by Dalenius and his colleagues, e.g.,
Bourke and Dalenius (1976) or Dalenius and Vitale (1979). From among the
recent papers about dealing with sensitive questions in population surveys,
we would like to mention, for example, papers by Trappmann et al. (2014)
and Kirchner (2015). In both of them, long lists of relevant references can
be found. Finally, recall that probably the most comprehensive account of
recent developments in sample survey theory and practice can be found in
Handbook of Statistics 29~A,~B, edited by Pfeffermann and Rao (2009).

\section{New randomized response technique}\label{newProcedure}

In this section, we suggest a completely different approach. Assume that
the studied sensitive variable~$Y$ is non-negative and bounded from above,
i.e., $0\leq Y\leq M$. Parameter~$M$ should be chosen taking into account
both bias and privacy. For that purpose, knowledge of the empirical
quantiles of the studied population, or at least reasonably guessing them,
is vital. Each respondent carries out, independently of the others, a
random experiment generating a pseudorandom number~$\Upsilon$ from the
uniform distribution on interval $(0,M)$, whereas the interviewer does not
know this value. The respondent can generate the pseudorandom
number~$\Upsilon$ using, for example, a laptop online/offline application;
for some other possibilities see sec.~\ref{randomMechanisms}. The
respondent then answers a simple question: ``\textit{Is the value of~$Y$ at
least~$\Upsilon$?}'' \big(e.g., ``\textit{Is your monthly income at
least~$\Upsilon$?}''\big). Note that the subscript~$AV$ used below
indicates that the estimator, as well as random variables used for its
construction, are based on the new idea of randomization suggested in this
Section.

Answer of the $i^{th}$ respondent follows the alternative distribution with
the parameter $Y_{i}/M$, i.e.,

\begin{equation}\label{ZiAV}
  Z_{i,AV,(0,M)} =
  \begin{cases}
    1\ \mathrm{with\ probability}\ \frac{Y_i}M,   &
       \mathrm{if}\ \Upsilon_{i}\leq Y_{i},\\
    0\ \mathrm{with\ probability}\ 1-\frac{Y_i}M, & \mathrm{otherwise}.
  \end{cases}
\end{equation}

\noindent
Evidently, $\E\big(Z_{i,AV,(0,M)}\big)= P\big(\Upsilon_{i}\leq
Y_{i}\big)=Y_i/M$ and
$\V\big(Z_{i,AV,(0,M)}\big)=\big(Y_i/M\big)\big(1-Y_i/M\big)$. If we
transform the answers $Z_{i,AV,(0,M)}$ to $R_{i,AV,(0,M)}=MZ_{i,AV,(0,M)}$,
then it holds

\begin{equation}\label{EvarRiAV}
  \E\big(R_{i,AV,(0,M)}\big) = Y_{i}
  \quad\mathrm{and}\quad
  \V\big(R_{i,AV,(0,M)}\big) = Y_{i}\big(M-Y_{i}\big).
\end{equation}

For certain sensitive variables, such as the total amount of alcohol
consumed within a certain period,  it is better to use a question:
``\textit{Is the value of~$Y$ lower than~$\Upsilon$?}'' In such a case we
recode the answer $Z_{i,AV,(0,M)}$ to
$Z_{i,AV,(0,M)}^{\star}=1-Z_{i,AV,(0,M)}$, and apply the suggested RRT to
$Z_{i,AV,(0,M)}^{\star}$.

\subsection{Application to the simple random sampling}
\label{application2SRS}

Consider now the situation in which the sampling plan $p(s)$ is a simple
random sampling without replacement with a fixed sample size~$n$. Denote by
$\overline{Y}=\tfrac1N\sum_{i \in U}Y_{i}$ the population mean, by
$S^2_{Y}=\tfrac1{N-1}\sum_{i \in U}\big(Y_{i}-\overline{Y}\big)^2$ the
population variance, and by $f=n/N$ the corresponding sampling fraction. In
this case, the inclusion probabilities are constant, i.e., $\pi_{i}=
P(\xi_i=1)=n/N\ \forall i\in U$.

Let the population total~$t_Y$ be estimated using the Horvitz-Thompson's
type estimator

\begin{equation}\label{nasOdhad}
  t_{HT,AV,(0,M)}^{R}\equiv t_{AV,(0,M)}^{R}=
  \frac{N}{n}\sum_{i \in s}R_{i,AV,(0,M)}.
\end{equation}

This estimator is evidently unbiased, and we calculate its variance. First,
taking into account the independence of outcomes of the randomization
experiments performed by the respondents, we have

\[
  \V{}_q\Big(t_{HT,AV,(0,M)}^{R}\,\big|\,s\Big)
 =\frac{N^2}{n^2}\sum_{i\in s}Y_i\big(M-Y_i\big).
\]

Using the well-known identity $\sum_{i \in U}\big(Y_{i}-\overline{Y}\big)^2
=\sum_{i \in U}Y^2_{i}-N\overline{Y}^2$, we can calculate the contribution
of the suggested RRT to the variance as

\begin{align*}
     \E{}_{p}\Big(\V{}_q\big(t_{HT,AV,(0,M)}^{R} \,\big|\, s\big)\Big)
  &=\E{}_{p}\Big(\frac{N^2}{n^2}\sum_{i\in s}Y_{i}\big(M-Y_{i}\big)\Big)
   =\E{}_{p}\Big(\frac{N^2}{n^2}\sum_{i\in U}Y_{i}\big(M-Y_{i}\big)\xi_i\Big)\\
  &=\frac{N}{n}\sum_{i\in U}Y_{i}\big(M-Y_{i}\big)
   =\frac{N^2}{n}\Big(\overline{Y}(M-\overline{Y})-\frac{N-1}{N}S^2_{Y}\Big).
\end{align*}

Finally, taking into account the variance of the simple random sampling
without replacement, see sec.~4.4 in Till\'{e} (2006) for details, we get

\begin{equation}\label{VtHTAVR}
  \V\big(t_{HT,AV,(0,M)}^{R}\big)=
  \frac{N^2}{n}\Big(\overline{Y}(M-\overline{Y})-\frac{n-1}{N}S^2_{Y}\Big).
\end{equation}

To characterize  the variance of the suggested estimators more profoundly,
and to get a more transparent insight into the variance of the suggested
RRT, we introduce two auxiliary ``measures of concentration''. More
precisely, let us denote

\begin{equation}\label{gym1}
   \Gamma_{Y,M}=\frac{1}{N} \sum_{i \in U}
   \frac{Y_{i}}{M}\Big(1-\frac{Y_{i}}{M}\Big)
  =\underbrace{\frac{1}{MN}\sum_{i \in U}
   Y_{i}}_{\frac{1}{M}\overline{Y}}-\underbrace{\frac{1}{M^2N}
   \sum_{i\in U} Y^2_{i}}_{\frac{1}{M^2}\overline{Y^2}}
  =\frac{\overline{Y}}{M}-\frac{\overline{Y^2}}{M^2}
\end{equation}
and
\begin{equation}\label{gym2}
  \Gamma_{\overline{Y},M}=\frac{\overline{Y}}{M}
  \frac{(M-\overline{Y})}{M}=\frac{\overline{Y}}
  {M}-\frac{\overline{Y}^2}{M^2}.
\end{equation}

\noindent
We call $\Gamma_{Y,M}$  \textit{the mean relative concentration measure},
and $\Gamma_{\overline{Y},M}$  \textit{the proximity measure of the
population mean $\overline{Y}$ to $\frac{M}{2}$}.

If $Y_{i}$ are i.i.d. random variables with a finite variance $\sigma^2$
and an expectation~$\mu$, then, by the law of large numbers, both
$\Gamma_{Y,M}$ and $\Gamma_{\overline{Y},M}$ converge, as $N\to\infty$,
with probability~1 to

\begin{equation}\label{gym3}
  \Gamma_{Y,M,as}=\frac{\mu}{M}\Big(1-\frac{\mu}{M}\Big)-\frac{\sigma^2}{M^2}
  \quad\mathrm{and}\quad
  \Gamma_{\overline{Y},M,as}=\frac{\mu}{M}\Big(1-\frac{\mu}{M}\Big).
\end{equation}

\noindent
We call $\Gamma_{Y,M,as}$ \textit{the asymptotic mean relative
concentration measure}, and $\Gamma_{\overline{Y},M,as}$ \textit{the
asymptotic proximity measure of the population mean $\overline{Y}$ to
$\frac{M}{2}$}. Note that both $\Gamma_{Y,M,as}$ and
$\Gamma_{\overline{Y},M,as}$ exist if $0\leq Y_{i}\leq M\ \forall i\in U$.

Let us focus on these measures in more detail. First, note that in our
setting both these measures are population characteristics, not random
variables. Second, both $\Gamma_{Y,M}$ and $\Gamma_{\overline{Y},M}$ take
on their values in the interval $[0,\tfrac{1}{4}]$, and are equal to zero
only in the pathological cases when either $Y_{i}=0\ \forall i\in U$ or
$Y_{i}=M\ \forall i\in U$. The higher these measures, the higher the
variance of $t_{HT,AV,(0,M)}^{R}$. The mean relative concentration
measure $\Gamma_{Y,M}$ attains its maximum $1/4$ when all values lie at the
center of the interval $(0,M)$, i.e., if $Y_{i}=M/2\ \forall i\in U$. The
measure $\Gamma_{\overline{Y},M}$ of the population mean's proximity to the
center of the interval $(0,M)$ attains its maximum $1/4$ only if the
population mean is at the interval center, i.e., $\overline{Y}=M/2$. This
case occurs, e.g., when random variable~$Y$ is symmetric around the center
of interval $M/2$; this feature is certainly true  for the uniform
distribution on $(0,M)$.

For a fixed value of the upper bound~$M$, population size~$N$ and sample
size~$n$, the contribution of the suggested RRT to the variance of
$t^{R}_{HT,AV,(0,M)}$ depends, up to a multiplicative constant, on
$\Gamma_{Y,M}$, because it holds

\begin{equation}\label{nasOdhadEp}
  \E{}_{p}\Big(\V{}_q\big(t_{HT,AV,(0,M)}^{R}\,\big|\,s\big)\Big)=
  \frac{M^2N^2}{n}
  \underbrace{\frac{1}{N}\sum_{i \in U}
  \frac{Y_{i}}{M}\Big(\frac{M-Y_{i}}{M}}_{\Gamma_{Y,M}}\Big)=
  \frac{M^2N^2}{n}\Gamma_{Y,M}.
\end{equation}

\noindent
Analogously, this contribution can  also be expressed, up to multiplicative
constants, by $\Gamma_{\overline{Y},M}$ and $S^2_Y$, because it holds

\begin{equation}\label{nasOdhadEpStar}
  \E{}_{p}\Big(\V\big(t_{HT,AV,(0,M)}^{R}\,\big|\,s\big)\Big)=
  \frac{M^2N^2}{n}\Gamma_{\overline{Y},M}-\frac{N(N-1)}{n}S^2_Y.
\end{equation}

Both $\Gamma_{Y,M}$ and $\Gamma_{\overline{Y},M}$ thus help us explain how
the suggested RRT increases the variance of the estimator of the population
total~$t_Y$ for  distributions symmetrical around $M/2$; for distributions
concentrated closely to the center of $(0,M)$, symmetrical around $M/2$; or
uniformly distributed. Moreover, they show that the suggested approach is
especially suitable for skewed distributions provided they are concentrated
around their mean values. Let us sum up: both these measures help us not
only describe the variance of the  estimator used, as well as
compare~\eqref{nasOdhadEp} and~\eqref{nasOdhadEpStar}, but also interpret
it better.

\smallskip\noindent
\textbf{Remark~1.}
Notice that, if the values of $Y$ are bounded both from below and above,
i.e., $0<m\le Y\leq M$, then variance of $t_{HT,AV,(0,M)}^{R}$ can be
significantly reduced by  generating pseudorandom numbers~$\Upsilon_{i}$
from the uniform distribution on the interval $(m,M)$ instead on $(0,M)$.
Indeed; if this is the case, we replace $Z_{i,AV,(0,M)}$, described
by~\eqref{ZiAV}, with

\[
  Z_{i,AV,(m,M)}=
  \begin{cases}
     1\ \mathrm{with\ probability}\ \frac{Y_i-m}{M-m},
     & m\le \Upsilon_{i}\leq Y_{i}, \\
     0\ \mathrm{with\ probability}\ 1-\frac{Y_i-m}{M-m},
     & \mathrm{otherwise},
  \end{cases}
\]

\noindent
transform these variables to $R_{i,AV,(m,M)}= m+(M-m)Z_{i,AV,(m,M)}$, and
estimate population total~$t_Y$ analogously to~\eqref{nasOdhad}, i.e.,
using the Horvitz-Thompson's type estimator

\begin{equation}\label{nasOdhadHvezda}
  t_{HT,AV,(m,M)}^{R}\equiv t_{AV,(m,M)}^{R}=
  \frac{N}{n}\sum_{i \in s}R_{i,AV,(m,M)}.
\end{equation}

\noindent
It is easy to show that the variance of $t_{HT,AV,(m,M)}^{R}$ is smaller
than that of $t_{HT,AV,(0,M)}^{R}$, namely, by the value
$\frac{N^2m}{n}\big(M-\overline{Y}\big)$.

When choosing parameters~$m$ and~$M$, both bias and privacy should be taken
into account. While the lower bound~$m$ affects mostly bias and is not
crucial for respondents' privacy, the choice of~$M$ affects both bias and
privacy. Thus, the knowledge of empirical quantiles for the studied
characteristic, or at least a reasonable guess about them, is vital for
setting~the values of~$m$ and~$M$ properly.

An immediate question arises of what happens if the interval $[m,M]$ has
not been set correctly. Evidently, if some values of $Y_{i}$ lie outside of
the  interval $[m,M]$, then with probability~1 it holds $Z_{i,AV,(m,M)}=0$
if $Y_{i}<m$ and $Z_{i,AV,(m,M)}=1$ if $Y_{i}>M$. The bias of the suggested
estimator then equals

\begin{equation}\label{vychyleni}
  \sum_{i \in U\,|\,Y_{i}<m} \big (Y_{i}-m\big) +
  \sum_{i \in U\,|\,Y_{i}>M} \big(Y_{i}-M\big)
\end{equation}

\noindent

\smallskip
Let us discuss some advantages of the suggested approach in comparison with
other currently used RRTs, including Eriksson's and Chaudhuri's:

\begin{itemize}
  \item It is simple; this fact  increases respondents'
  confidence and
        cooperation, and thus reduces the estimation error.

  \item Respondents' privacy is well protected, because they never report
        the true value of the sensitive variable.

  \item One can avoid a demanding task of designing the deck of cards
        to  mask the studied variable.

  \item It enables us to estimate the population total at an acceptable
        level of accuracy, see sec.~\ref{simulations} for details.

\end{itemize}

\noindent
Due to the device/technique used for generating random numbers, some
respondents may feel a lower degree of  confidence in preserving their
anonymity.

A natural question arises whether we could improve the accuracy of the
suggested method.  We discuss two modifications of the RRTs suggested above
and their properties in the subsections below. The heuristics behind this
approach are based on the following observations. All the techniques
presented up to now have assumed that the interviewer does not know the
outcome of the random mechanism leading to the randomized response, such as
the card drawn, the value of the pseudorandom number, etc. It is plausible
to ask what would happen if we also knew  the outcome of that random
experiment on the one hand, while protecting respondents' privacy on the
other hand. More precisely: can statisticians improve the accuracy of the
proposed estimator, i.e., to decrease its variance, if they  also know the
values of the generated pseudorandom numbers? We surmise it is feasible,
and suggest one possible way of reaching this goal. Let us point out,
however, that the success of the suggested approach, to a considerable
extent, depends on the statistician's insight into  the problem. It may be
embarrassing to report either high or  low values of the variables in
question, say, the personal income. Depending on the value of the
pseudorandom number~$\Upsilon$, a different question is then asked with the
aim to reduce the respondent's potential embarrassment.

\subsection{Estimators using knowledge of~$\Upsilon$}
\label{knowledgeUpsilon}

Assume again that the studied sensitive variable~$Y$ is non-negative and
bounded from above, i.e., $0\leq Y\leq M$. Each respondent carries out,
independently of the others, a random experiment generating a pseudorandom
number~$\Upsilon$ from the uniform distribution on interval $(0,M)$, and
\textit{informs the interviewer of both its value} and \textit{whether
$\Upsilon\leq Y$ or not}. For example, the response is that the simulated
number has been~$xxx$ and the respondent earns more/less. Assume further
that the corresponding random response is now described not
by~\eqref{ZiAV}, but using a dichotomous random variable

\begin{equation}\label{ZiAValfa}
 Z_{i,AV,\alpha} =
  \begin{cases}
    1-\alpha + 2\alpha\frac{\Upsilon_{i}}{M}, &
      \mathrm{if}\ \Upsilon_{i}\leq Y_{i},\\
     -\alpha + 2\alpha\frac{\Upsilon_{i}}{M}, &\mathrm{otherwise,}
   \end{cases}
   \qquad 0\leq\alpha<1.
\end{equation}

For random responses $Z_{i,AV,\alpha}$ it holds

\[
  \E\big(Z_{i,AV,\alpha}\big)=P\big(\Upsilon_i\leq Y_i\big)=
  \frac1M\int_0^{Y_i} \Big(1-\alpha+2\alpha\frac{u}{M}\Big)\,du
  +\frac1M\int_{Y_i}^M \Big(-\alpha+2\alpha\frac{u}{M}\Big)\,du
  =\frac{Y_{i}}{M},
\]
\[
  \V\big(Z_{i,AV,\alpha}\big)
  =\frac{1-2\alpha}{M^2}Y_{i}\big(M-Y_{i}\big)+\frac{\alpha^2}{3}.
\]

\noindent
The random responses $Z_{i,AV,\alpha}$ are transformed to
$R_{i,AV,\alpha}=MZ_{i,AV,\alpha}$, and the desired estimator of the
population total~$t_Y$ can be constructed analogously to~\eqref{nasOdhad}
and~\eqref{nasOdhadHvezda}. More precisely, we suggest using again the
Horvitz-Thompson's type of estimator in the form

\begin{equation}\label{vozar1}
  t_{HT,AV,\alpha}^R \equiv t_{AV,\alpha}^R =
  \frac{N}{n} \sum_{i \in s} R_{i,AV,\alpha}.
\end{equation}

Because $\E\big(R_{i,AV,\alpha}\big)=Y_i$, estimator~\eqref{vozar1} is
unbiased, and the contribution of the randomization to its variance is

\begin{equation}\label{varComp}
  \E{}_{p}\Big(\V{}_q\big(t_{HT,AV,\alpha}^R\,\big|\,s\big)\Big)
  =\frac{M^2N^2}{n} \sum_{i\in U}
  \Big(\frac{1}{N}\big(1-2\alpha\big)\frac{Y_{i}}{M}
  \Big(1-\frac{Y_{i}}{M}\Big)+\frac{\alpha^2}{3N}\Big).
\end{equation}

\noindent
An easy calculation shows that \eqref{varComp} takes on its global minimum
at  $\alpha_{opt}=3\Gamma_{Y,M}\in[0,3/4]$. Substituting $\alpha_{opt}$
back to~\eqref{varComp}, we get

\begin{align}\label{EptR}
   \E{}_{p}\Big(\V{}_q\big(t_{HT,AV,\alpha_{opt}}^{R}\,\big|\,s\big)\Big)
  &=\frac{M^2N^2}{n}\sum_{i \in U}\Big(\big(1-6\Gamma_{Y,M}\big)
   \frac{1}{N}\frac{Y_{i}}{M}\Big(1-\frac{Y_{i}}{M}\Big)+
   \frac{3\Gamma_{Y,M}^2}{N}\Big) \notag \\
  &=\frac{M^2N^2}{n}\Gamma_{Y,M}\big(1-3\Gamma_{Y,M}\big).
\end{align}

We would like to point out that the knowledge of pseudorandom
numbers~$\Upsilon_{i}$ and the use of $\alpha_{opt}$ can considerably
decrease variability depnding on the suggested RRT -- compare~\eqref{EptR}
with~\eqref{nasOdhadEp}. Note also that our simulations summarized in
sec.~\ref{simulations} confirm these findings.

The value of the parameter~$\alpha$, which is a priori set by the
interviewer, is fixed and unknown to the respondent. For $\alpha=0$ we have
the original method described in sec.~\ref{application2SRS}.  The response
to $Z_{i,AV,\alpha}$ is transformed not by the respondent, but by the
interviewer off-line.

Parameter~$\alpha$  should be set to its optimal value
$\alpha_{opt}=3\Gamma_{Y,M}$, where the mean relative concentration measure
$\Gamma_{Y,M}$ is introduced in sec.~\ref{newProcedure},
formula~\eqref{gym1}. If the interviewer has some prior information about
the mean~$\mu$ and variance~$\sigma^{2}$ values for the theoretical
distribution of the surveyed variable~$Y$, he/she should rather apply
asymptotic concentration measure~\eqref{gym3}, which can be estimated using
a plug-in moment estimator. More precisely, the population
mean~$\overline{Y}$ should be replaced with~$\mu$, and the population
variance~$S^{2}_{Y}$ with~$\sigma^{2}$. Since the population second moment
$\overline{Y^{2}}$ can be expressed as $\frac{N-1}N
S^{2}_{Y}+\overline{Y}^2$, it is sufficient to substitute
$\mu$~and~$\sigma^{2}$ into this expression. Recall that the prior
information is often available for regular surveys in official statistics,
such as EU-SILC, because in such a case we can either use results from
previous years updated by inflation, or we can rely on the expert opinion.
If no prior information is available, we recommend  choosing small values
of $\alpha$, such as $0.5$, to decrease the negative values of
$R_{i,AV,\alpha}$. Note that in such a case the resulting estimator may
attain unacceptably low or even negative values for estimates of
non-negative variables. However this issue can, to some extent, be resolved
by properly tuning parameter~$\alpha$ and increasing the sample size.

Notice that if a non-negative surveyed random variable~$Y$ is bounded not
only from above, but also from below, i.e., $0<m\leq Y\leq M$, we generate
$\Upsilon_i$ from the uniform distribution on the interval $(m,M)$,
modifying $Z_{i,AV,\alpha}$ given by~\eqref{ZiAValfa} to

\[
  Z_{i,AV,\alpha,(m,M)} =
  \begin{cases}
     1-\alpha+2\alpha\frac{\Upsilon_{i}}{M-m}, &
       \mathrm{if}\ \Upsilon_{i}\leq Y_{i},\\
      -\alpha+2\alpha\frac{\Upsilon_{i}}{M-m}, & \textrm{otherwise,}
  \end{cases}
  \qquad 0\le\alpha<1,
\]

\noindent
transforming $Z_{i,AV,\alpha,(m,M)}$ to $R_{i,AV,\alpha,(m,M)}=
(M-m)Z_{i,AV,\alpha,(m,M)} +m(1-2\alpha)$, and forming an estimator of the
population total~$t_{Y}$ of the Horvitz-Thompson's type analogously
to~\eqref{vozar1}, i.e.,

\begin{equation}\label{vozar2}
  t_{HT,AV,\alpha,(m,M)}^R \equiv t_{AV,\alpha,(m,M)}^R =
  \frac{N}{n} \sum_{i \in s} R_{i,AV,\alpha,(m,M)}.
\end{equation}

\noindent
Because $\E\big(R_{i,AV,\alpha},(m,M)\big)=Y_i$, the
estimate~\eqref{vozar2} is again unbiased.

\zapomen{MATHEMATICA CONTROL
 Integrate[(1/M) (1 - al + 2 al u/M), {u, 0, Y}] +
 Integrate[(1/M) (-al + 2 al u/M), {u, Y, M}]
 gives Y/M
 Simplify[
 Integrate[(1/(M - m)) (1 - al + 2 al u/(M - m)), {u, m, Y}] +
  Integrate[(1/(M - m)) (-al + 2 al u/(M - m)), {u, Y, M}]]
 gives Y - m(1-2al)
 Simplify[
 Integrate[(1/M) (1 - al + 2 al u/M)^2, {u, 0, Y}] +
  Integrate[(1/M) (-al + 2 al u/M)^2, {u, Y,
    M}] - (Integrate[(1/M) (1 - al + 2 al u/M), {u, 0, Y}] +
     Integrate[(1/M) (-al + 2 al u/M), {u, Y, M}])^2]
 gives
 al^2/3 + ((M - Y) Y)/M^2 + (2 al Y (-M + Y))/M^2
 so that
 FullSimplify[(1 - 2 al) Y (M - Y) - (1 - 2 al) Y (M - Y)]
 gives
 0
 and also
 Simplify[
 Integrate[(1/M) (1 - al + 2 al u/M)^2, {u, 0, Y}] +
  Integrate[(1/M) (-al + 2 al u/M)^2, {u, Y,
    M}] - (Integrate[(1/M) (1 - al + 2 al u/M), {u, 0, Y}] +
     Integrate[(1/M) (-al + 2 al u/M), {u, Y, M}])^2 -
  al^2/3 - ((1 - 2 al) Y (M - Y))/M^2]
 gives
 0}

We must firmly emphasize here that the information about neither the value
of pseudorandom number~$\Upsilon$ nor of the value ~$\alpha$ enables us to
guess the exact value of the sensitive variable~$Y$, except for the case
$Y=M$. In other words, knowing them  does not intrude on the respondent's
privacy.

The heuristics behind the proposed modification are the following:

\begin{itemize}
  \item If the answer is \textit{YES}, then a high value of the
        pseudorandom number $\Upsilon$ implies a high value of the studied
        variable $Y$, because $Y\geq \Upsilon$, and these observations
        ``considerably'' increase the value of the estimator.

  \item On the other hand, if the answer is \textit{NO}, then a low value
        of the pseudorandom number $\Upsilon$ implies a low value of $Y$,
        because $Y<\Upsilon$, and these observations ``considerably''
        decrease the value of the estimator.

\end{itemize}

\noindent
Unfortunately, in both of these situations, i.e., when the value of the
response is either (too) low or (too) high, the respondent may be more
prone to fabricate his/her answer.

\subsection{Estimators using switching questions}
\label{switchingQuestions}

Let us emphasize that for some characteristics, such as monthly income of a
household, it may be sensitive for respondents to report either high or low
values. This led us to modifying the suggested RRT approach in the
following way.

First, we set a proper fixed threshold~$T$, $0<T<M$, unknown to the
respondent. Depending on whether the pseudorandom number~$\Upsilon$, which
is distributed according to the uniform distribution on $(0,M)$, does or
does not exceed the fixed threshold~$T$, we ask one of the following
questions:

\begin{enumerate}
\renewcommand{\theenumi}{\roman{enumi}}
  \item If $\Upsilon \leq T$: ``\textit{Is the value of~$Y$ at least~$\Upsilon$?}'',
  \item If $\Upsilon>T$: ``\textit{Is the value of~$Y$ smaller than~$\Upsilon$?}''.
\end{enumerate}

Second, we form random variables
\begin{equation}\label{vozar3}
  Z_{i,AV,T} =
  \begin{cases}
   \phantom{-}1, &\mathrm{if}\  \Upsilon_{i} \leq T, \ \Upsilon_{i} \leq Y_{i}, \\
   \phantom{-}0, &\mathrm{if}\  \Upsilon_{i} \leq T, \ \Upsilon_{i} > Y_{i},
     \ \mathrm{or\ if }\ \ \Upsilon_{i} > T, \ \Upsilon_{i} \leq Y_{i}, \\
   -1,           &\mathrm{if}\ \Upsilon_{i} > T,    \ \Upsilon_{i} >  Y_{i}.   \\
  \end{cases}
\end{equation}

If we know only the answer concerning the value of~$Y$ but not the question
asked, i.e., whether $\Upsilon_{i}\leq T$ or not, $Z_{i,AV,T}$ has the
expectation $\E\big(Z_{i,AV,T}\big)=1- |\,T/M-Y_{i}/M\,|$. Unfortunately,
in such a case it is impossible to construct either an estimator of the
population total~$t_Y$ or of the population mean~$\overline{t}_Y$.

On the other hand, if we know both the answer concerning the value of~$Y$
and the question asked, i.e., whether $\Upsilon_{i} \leq T$ or not, then
$\E\Big(Z_{i,AV,T}\Big)=Y_{i}/M+T/M-1$. In this case, the  transformation
of $Z_{i,AV,T}$ to $R_{i,AV,T}=MZ_{i,AV,T}+M-T$ enables us to construct an
unbiased estimator of the population total~$t_Y$ of the Horvitz-Thompson's
type, which has the form

\begin{equation}\label{vozar1T}
  t_{HT,AV,T}^R\equiv t_{AV,T}^R=\frac{N}{n} \sum_{i \in s} R_{i,AV,T}.
\end{equation}

\noindent
An unbiased estimator of the population mean~$\overline{t}_Y$ can be
constructed analogously.

As regards the variance of $R_{i,AV,T}$, we must distinguish between
$Y_{i}>T$ and the complementary inequality. It holds

\begin{equation}\label{rozptylZ}
  \V\big(R_{i,AV,T}\big) =
  \begin{cases}
    Y_{i}\big(M-Y_{i}\big)+(M-T)\big(2Y_{i}+T\big), &\ Y_{i} \leq T, \\
    Y_{i}\big(M-Y_{i}\big)+T\big(3M-2Y_{i}-T\big),  &\ Y_{i} > T.    \\
 \end{cases}
\end{equation}

\noindent
If we compare~\eqref{rozptylZ} with~\eqref{EvarRiAV}, we can see that the
variance of~$R_{i,AV,T}$ is always higher than that of $R_{i,AV,(0,M)}$.
Because negative values of $Z_{i,AV,T}$ may occur, this may  occasionally
lead to negative values of $R_{i,AV,T}$. More precisely, note that
$R_{i,AV,T}$ attains only three values, i.e., positive (equal to~$2M-T$),
zero, and negative (equal to~$-T$), being the source of its poor
performance. Recall that $t_{HT,AV,T}^R$ is intended to estimate
non-negative variable~$Y$. Unfortunately, looking at the results of our
simulations we observe that $t_{HT,AV,T}^R$  quite often returns
inadmissibly low, or even negative values; this is a big drawback.

A simple, but somewhat tedious, analysis of~\eqref{rozptylZ} shows that we
cannot find the optimal value of the threshold from an open interval
$0<T<M$ minimizing $\V\big(R_{i,AV,T}\big)$. Moreover, numerical
experiments show that the variance of $\V\big(R_{i,AV,T}\big)$ is
acceptable only for very low, or very high, values of the threshold~$T$,
like $T=0.1$ or $T=0.9M$. For example, the variance contribution of the
modified RRT for $T=0.9M$ is

\begin{align*}
  \E{}_{p}\Big(\V{}_q\big(t_{HT,AV,T=0.9M}^{R}\big| s\big)\Big)=
  \frac{N}{n}\Bigg(&\sum_{i \in U} Y_{i}\big(M-Y_{i}\big)
 +\sum_{i \in U\,|\,Y_{i}\le T}\big(0.20Y_{i}+0.09M\big) \\
  &+\sum_{i \in U\,|\,Y_{i} >T}\big(-1.80Y_{i}+1.89M\big)\Bigg).
\end{align*}

Notice that if a non-negative surveyed random variable~$Y$ is bounded not
only from above, but also from below, i.e., $0<m\leq Y\leq M$, we
generate~$\Upsilon$ from the uniform distribution on $(m,M)$, and,
analogously to~\eqref{vozar3}, form random variables

\[
  Z_{i,AV,T,(m,M)} =
  \begin{cases}
   \phantom{-}1, &\mathrm{if}\ \Upsilon_{i}\leq T,\Upsilon_{i}\leq Y_{i},\\
   \phantom{-}0, &\mathrm{if}\ \Upsilon_{i}\leq T,\Upsilon_{i}>Y_{i},
      \ \mathrm{or\ if}\ \ \Upsilon_{i}>T,    \Upsilon_{i}\leq Y_{i},\\
   -1,           &\mathrm{if}\ \Upsilon_{i}>T,    \Upsilon_{i}>Y_{i}.
  \end{cases}
\]
\zapomen{MATHEMATICA CONTROL
  Simplify[
  Integrate[1/(M - m), {u, m, T}] - Integrate[1/(M - m), {u, Y, M}]]
  gives (m+M-T-Y)/(m-M)
  Simplify[m + M - T + (M - m) Simplify[
    Integrate[1/(M - m), {u, m, T}] - Integrate[1/(M - m), {u, Y, M}]]]
  gives Y}

\noindent
It is easy to show that $\E\big(Z_{i,AV,T,(m,M)}\big)=(T+Y_i-m-M)/(M-m)$,
so that if we transform $Z_{i,AV,T,(m,M)}$ to $R_{i,AV,T,(m,M)}
=(M-m)Z_{i,AV,T,(m,M)} +m+M-T$, then $\E\big(R_{i,AV,T,(m,M)}\big)=Y_i$.
Now we can form unbiased estimator of the population total~$t_{Y}$ of the
Horvitz-Thompson's type, analogously to~\eqref{vozar1}, of the form

\begin{equation}\label{vozar2T}
  t_{HT,AV,T,(m,M)}^R\equiv t_{AV,T,(m,M)}^R=\frac{N}{n} \sum_{i \in s}
  R_{i,AV,T,(m,M)}.
\end{equation}

Let us point out that modifications described in this Section are
interesting especially from the theoretical point of view. Despite them
offering a seemingly nice idea, they cannot be recommended for practical
use. We can also compare the results of the simulations.

\subsection{Random number generation}\label{randomMechanisms}

In all RRTs we are aware of, the preparation of the random mechanism is
probably the trickiest point. For example, it is not clear how to design an
acceptably large deck of cards that would sufficiently mask the true values
(interviewer cannot guess very close to the true values using respondents'
answers and the knowledge of cards from this deck) and provide sufficient
accuracy. Assume now direct face-to-face interviewing and describe several
possibilities for generating random numbers.

\begin{enumerate}
  \item We allow the respondent to select the random number according to
        the European  ISO 28640:2010(en) Standard, which provides not only
        the methods suitable for generation, but also tables of random
        numbers and random digits. Recall that equivalents of this
        Standard, as well as of the tables of random numbers, exist all
        over the world. We are convinced that existence of an international
        standard can increase credibility of the survey and willingness of
        respondents to respond truthfully. The selected random number is
        then used according to the RRT used.

  \item To those who feel they are ``experts in the field of randomness'',
        the reviewer can offer that they select a random number from the
        uniform distribution using his/her own method. The remaining
        procedure is the same as described above.

  \item Another possibility is, e.g., using a huge deck of cards, for
        example cards with 100-CZK value steps in our case, but it would
        require additional  calculations to find the bias of such an
        approach.

\end{enumerate}

On the other hand, we would like to point out that the question of
credibility is not only a matter for statisticians, but more and more a
task for psychologists. While statisticians must suggest procedures which
are ``sufficiently random'' in their eyes, psychologists must find and
offer ways to convince the respondents that they are not cheated.
Unfortunately, a detailed discussion of this topic would go beyond the
scope of this paper.

\section{Simulation Study}\label{simulations}

In many countries, income is recognized as a private and (highly) sensitive
item of information. The respondents often refuse to respond at all or
provide strongly biased answers. This in particular happens if their income
is (very) high or (very) low. That leads us to assessing the performance of
the proposed RRT by a simulation study using Czech wage data from the
Average Earnings Information System (IPSV) of the Ministry of Labor and
Social Affairs of the Czech Republic.

Based on the extensive analysis of monthly wage data provided by IPSV from
the years 2004\,--\,2014, Vrabec and Marek (2016) recommended a model of
wages in the Czech Republic as a three-parameter log-logistic distribution
with the density

\begin{equation}\label{densityLogLogistic}
 f(y;\tau,\sigma,\delta) =
  \begin{cases}
   \frac{\tau}{\sigma}\big(\frac{y-\delta}{\sigma}\big)^{\tau-1}
   \Big(1+\big(\frac{y-\delta}{\sigma}\big)^{\tau}\Big)^{-2}, &
   y\geq\delta>0,\ \tau>0,\ \sigma>0, \\
    0,       & \textrm{otherwise,}
  \end{cases}
\end{equation}

\noindent
where~$\tau>0$  is a shape parameter, $\sigma>0$~is a scale parameter, and
$\delta$ is a location parameter.

We estimate parameters of~\eqref{densityLogLogistic} using the data from
$2^{nd}$ quarter 2014, and receive
\begin{equation}\label{odhady}
  \widehat\tau=4.0379,\ \widehat\sigma=21,687\ \ \textrm{and}
  \ \ \widehat\delta=250.
\end{equation}

\noindent
The corresponding estimated average monthly income is $24,290$~CZK
(approximately $950$ EUR). Note that the estimates~\eqref{odhady} are based
on roughly $2.1\times10^6$ observations, covering practically half of the
overall relevant population.

\begin{figure}[H]
\begin{center}
  \includegraphics[scale=0.87]{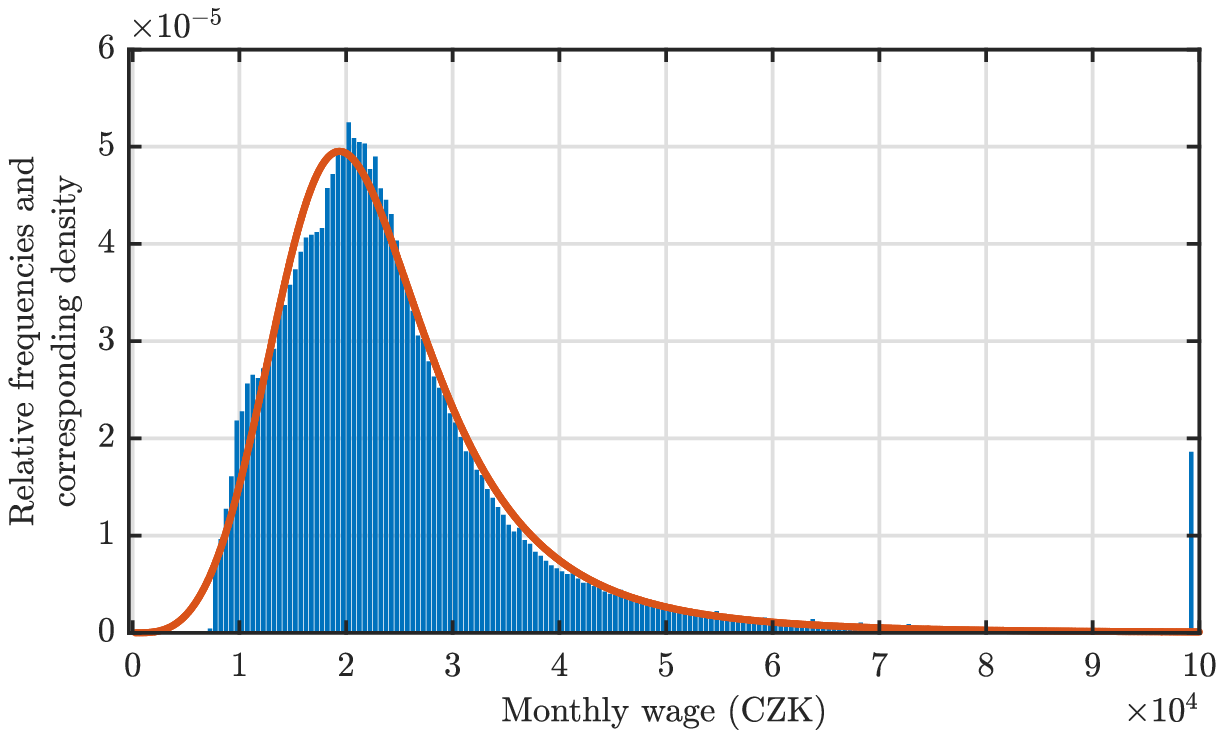}
  \caption{Probability histogram of monthly wages in the Czech Republic in
           the 2nd quarter of 2014, and the density (in red) of
           approximating model~\eqref{densityLogLogistic} with the
           parameters estimated by~\eqref{odhady}.}
  \label{histogram}
\end{center}
\end{figure}

\begin{figure}[H]
\begin{center}
  \includegraphics[scale=0.87]{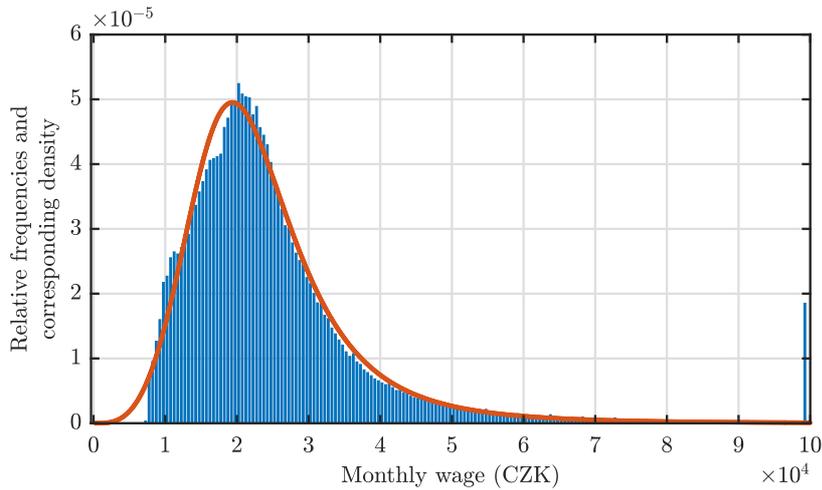}
  \caption{The sample quantile function of monthly wages in the Czech
           Republic in the 2nd quarter of 2014.}
  \label{kvantily}
\end{center}
\end{figure}

Histograms of the data with the bin width 500 (CZK), and density of the
log-logistic distribution~\eqref{densityLogLogistic} with the unknown
parameters replaced by their estimates~\eqref{odhady}, are presented in
fig.~\ref{histogram}. Moreover, the corresponding sample quantile function
of the observed wages is presented in fig.~\ref{kvantily}. It is
interesting to take a look at both lower and upper sample quantiles of the
data used. While $8,000$~CZK corresponds to the $0.01$~sample quantile,
$40,000$~CZK corresponds to the $0.91$~sample quantile, $60,000$~CZK to the
$0.97$~sample quantile and, finally, $80,000$~CZK to the $0.98$~sample
quantile, compare visually fig.~\ref{kvantily}.

It is evident from fig.~\ref{histogram} that the original data is highly
skewed. Therefore, it is not surprising that the mean relative
concentration measure $\Gamma_{Y,M}=0.198$ is close to its attainable
maximum. In such a case, as follows from sec.~\ref{newProcedure}, we can
expect higher variance of the estimators using the suggested RRT than for
the Horvitz-Thompson's estimator based on non-randomized data. Moreover,
the estimator $t^R_{HT,AV,\alpha}$ based on the knowledge of
$\Upsilon_{i}$'s and ``almost-optimal'' choice of the parameter
$\alpha\approx3\Gamma_{Y,M}$, should have smaller variance than
$t^R_{HT,AV,(m,M)}$ (corresponding to $\alpha=0$). That conjecture is
confirmed by our simulations.

From the ``theoretical'' wage distribution corresponding to
model~\eqref{densityLogLogistic}, in which unknown parameters have been
replaced with their estimates~\eqref{odhady}, 1,000
replications\footnote{Note that the simulation results virtually do not
change after 100 replications of the population; differences begin at the
third significant digit.} of populations sized $N=200$, or $N=400$ are
simulated. The simulations are carried out with the aid of statistical
freeware~R, version 3.5.1; for details, see R~Core Team (2018). Data from
the log-logistic distribution is generated using the package
\textit{flexsurv}.

From each replication of the population, we draw, without replacement,
1,000 random samples of the size $n=20$,\ or $n=50$. Such population and
sample sizes are standard for separate strata in the business sampling
surveys, and  also resemble the usual social statistical surveys, such as
the EU Statistics of Income Living Condition. In such a survey for a medium
sized country like the Czech Republic with the population of 10,000,000
inhabitants and approximately $4,300,000$ households, the samples
approximately include $9,500$ households surveyed in a two-dimensional
stratification (region and size of municipality), giving $78\times4=312$
strata. The average sample size is then about ~$30$ per stratum. In
EU-SILC,  detailed results are presented for eight income groups, leading
on average to the population size of approximately $N=1,250,000$
inhabitants per one income group. For a more detailed description of the
stratification, strata, sample sizes and sampling design see
\textit{EU-SILC 2016}.

For each sample, both $t_Y$ and $\overline t_{Y}$ are estimated using the
techniques described in sec.~\ref{newProcedure}. Estimates of the total
mean values, instead of the population totals, are presented to enable
easier comparison between the results obtained for populations with
different sizes~$N$ and different sample sizes~$n$.

In the simulations, we are especially interested in the impact of ``tuning
parameters'' $m, M, T, \alpha$ and $\alpha_{opt}$ on the estimates. Taking
into account the type and nature of the data we are simulating, we set the
parameters as described in tab.~\ref{parametry}. The values
of~$\alpha_{opt}$ were set using the formulae for the optimal variance
described in sec.~\ref{newProcedure}. Other parameters were chosen with
regard to our experience, in particular, which monthly salary can be
perceived to be high. Because practically all the available data is larger
than 7,000 CZK, we set the lower bound of the interval for generating
pseudorandom numbers~$\Upsilon_i$ to $m=7,000$.

\smallskip
\begin{table}[H]
\begin{center}
  \setlength{\extrarowheight}{5pt}
  \begin{tabular}{|r|r|r|r|r|}
  \hline
        $m$    &       $M$ &       $T$ & $\alpha$ & $\alpha_{opt}$  \\
  \hline
     $7,000$  & $40,000$ & $30,000$ &   $0.75$ &         $0.72$  \\
     $7,000$  & $60,000$ & $45,000$ &   $0.75$ &         $0.59$  \\
     $7,000$  & $80,000$ & $45,000$ &   $0.75$ &         $0.52$  \\
  \hline
  \end{tabular}
  \medskip
  \caption{Choice of tuning parameters for the simulations.}
          \label{parametry}
\end{center}
\end{table}

\begin{figure}[H]
\begin{center}
  \includegraphics[scale=1]{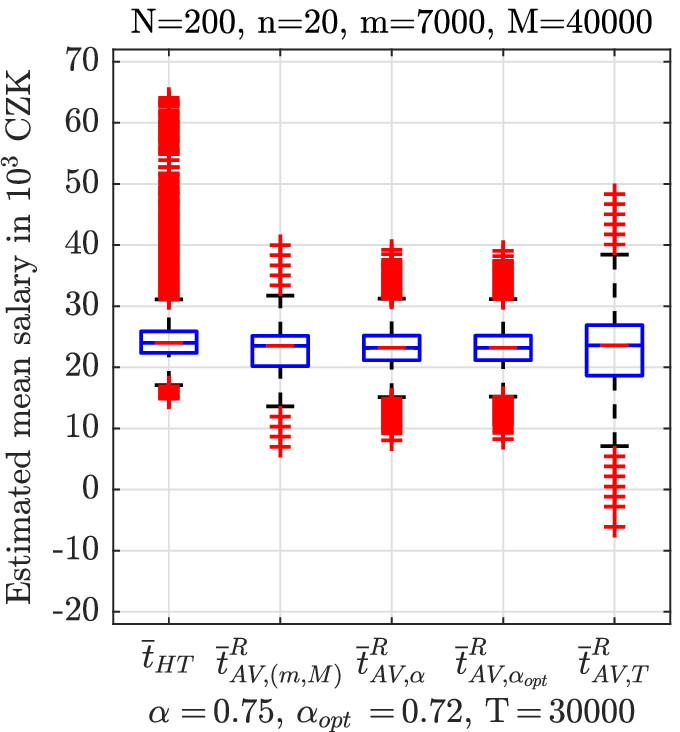}
  \includegraphics[scale=1]{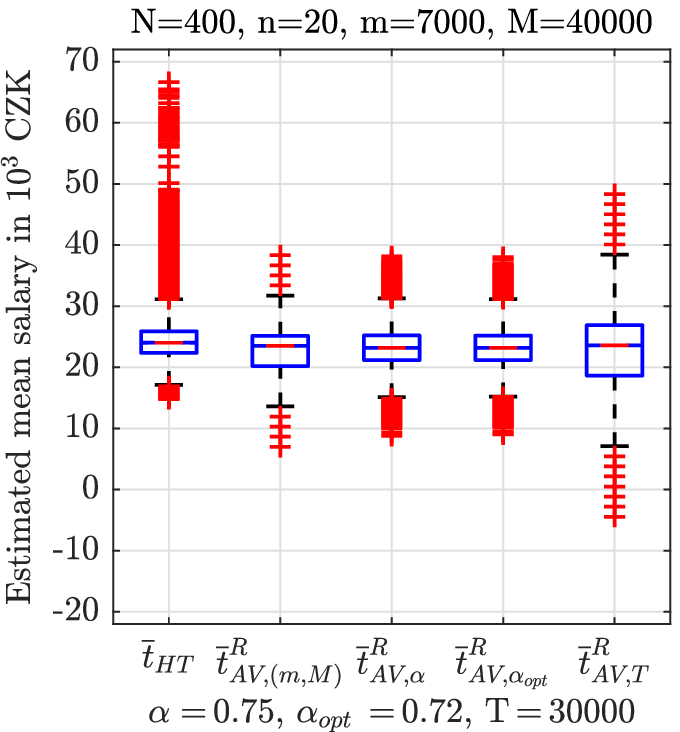}

  \bigskip
  \includegraphics[scale=1]{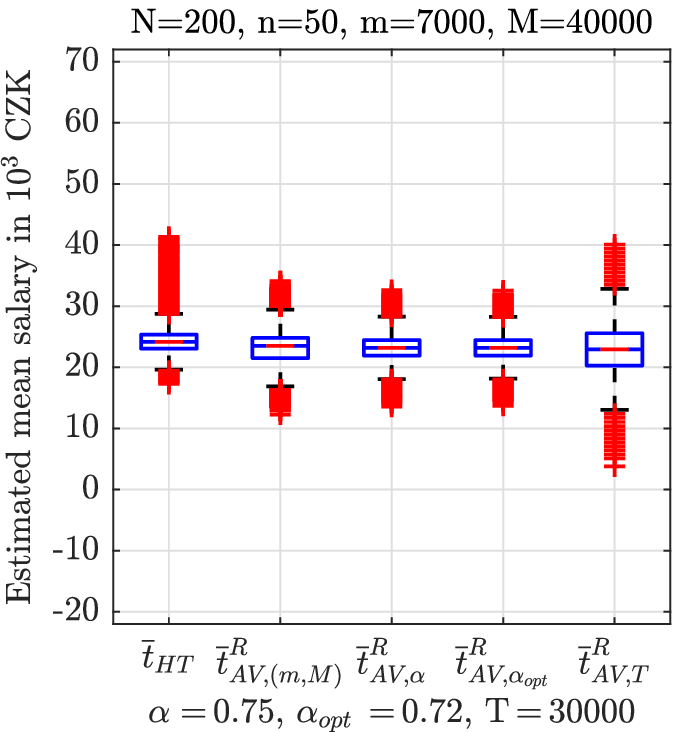}
  \includegraphics[scale=1]{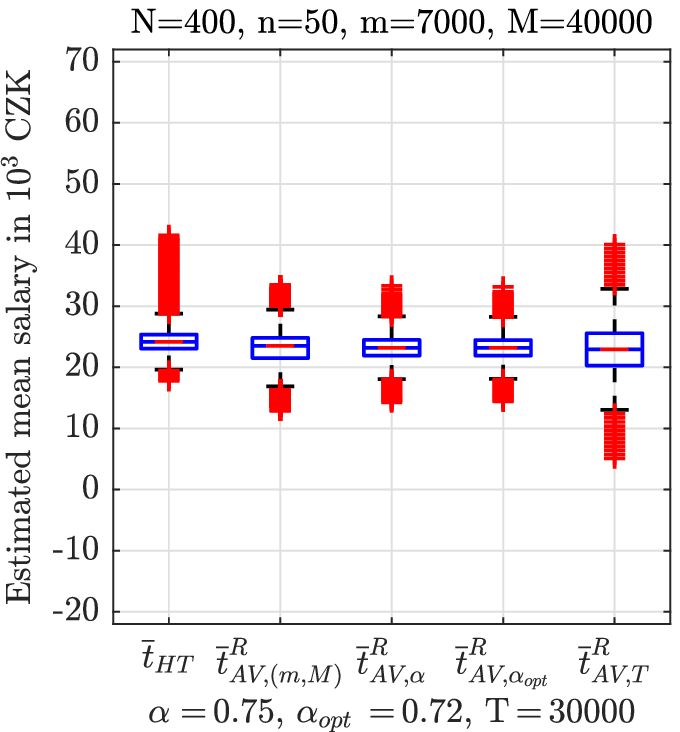}

  \caption{Behavior  of considered estimators applied to different sizes of
           the population~$N$ and sample sizes~$n$; $(m,M)=(7,000;
           40,000)$, $T=30,000$, $\alpha=0.75$ and $\alpha_{opt}=0.72$.}
  \label{results40}
\end{center}
\end{figure}

\begin{figure}[H]
\begin{center}
  \includegraphics[scale=1]{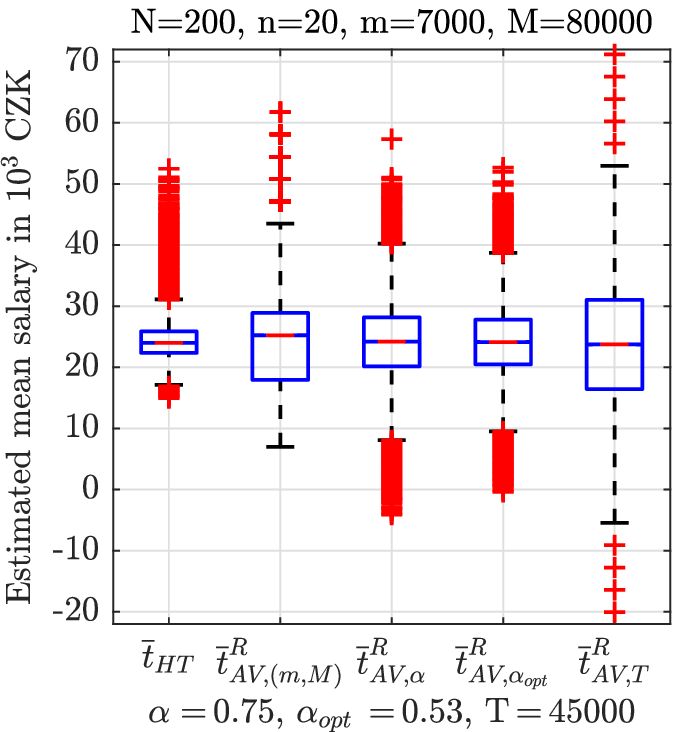}
  \includegraphics[scale=1]{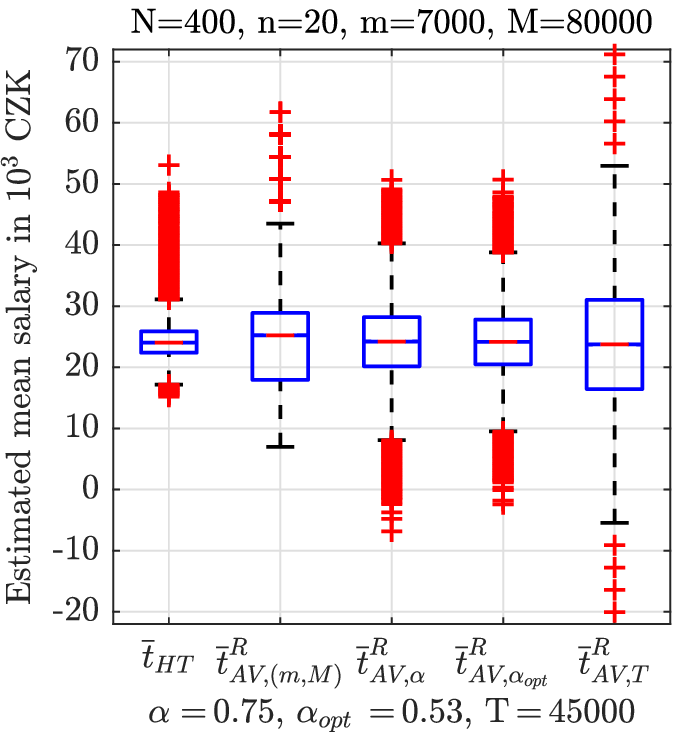}

  \bigskip
  \includegraphics[scale=1]{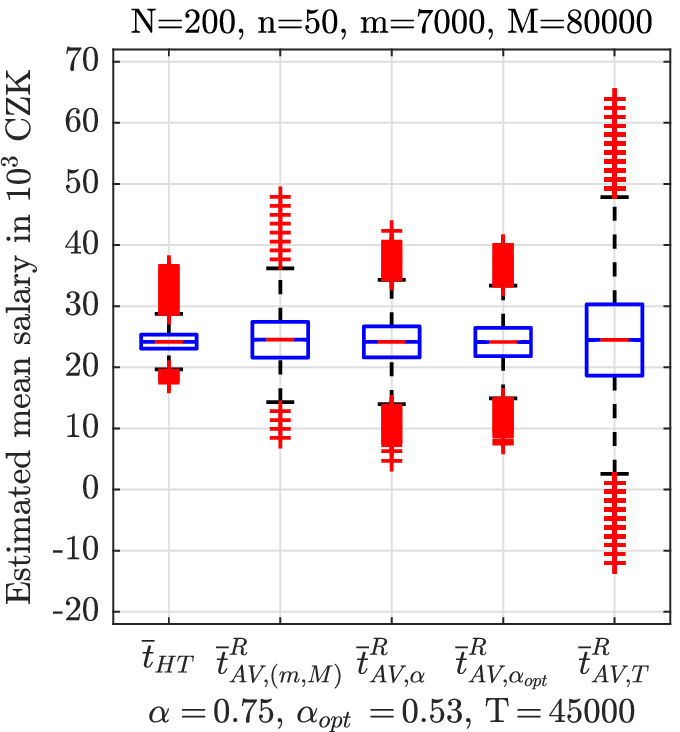}
  \includegraphics[scale=1]{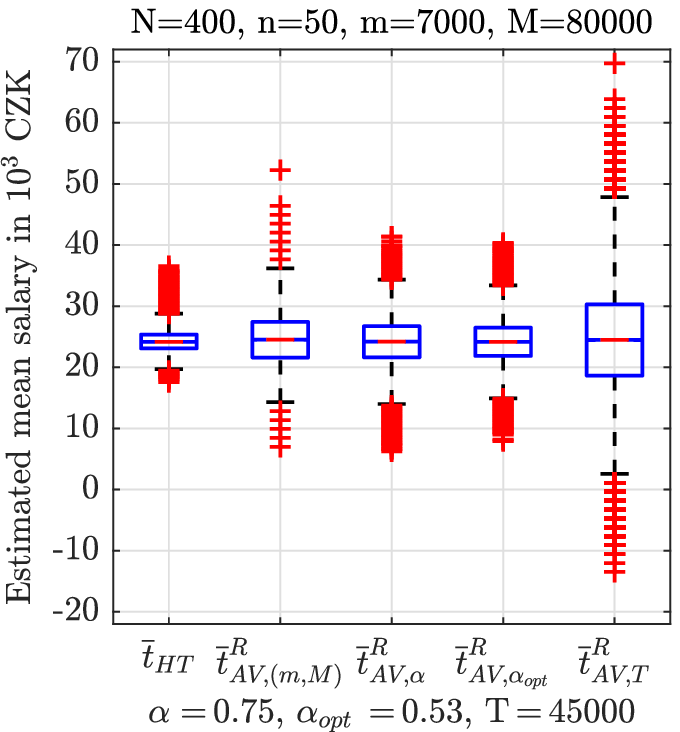}
  \caption{Behavior  of considered estimators applied to different sizes of
           the population~$N$ and sample sizes~$n$; $(m,M)=(7,000;
           80,000)$, $T=45,000$, $\alpha=0.75$ and $\alpha_{opt}=0.53$.}
  \label{results80}
\end{center}
\end{figure}

\smallskip
The results are summarized
 \footnote{In tab.~\ref{tabRes1}\,--\,\ref{tabRes3} both the sample
  averages (mean) and sample standard deviations (sd) of the simulated
  values are presented. For simplicity, we omit ``HT'' in the descriptions
  of the analyzed estimators in all figures and tables because all the
  estimators we compare here are of the Horvitz\,--\,Thompson's type.}
in tab.~\ref{tabRes1}\,--\,\ref{tabRes3}
and in fig.~\ref{results40}\,--\,\ref{results80}. They show that for large
populations the accuracy of the suggested estimators is acceptable even for
the method of the switching questions. The reason for the lower standard
deviation of $\overline t_{AV,\alpha}^R$, and especially $\overline
t_{AV,\alpha_{opt}}^R$, in comparison with $\overline t_{AV,T}^R$ and
$\overline t_{AV,(m,M)}^R$ is that this estimator efficiently uses the
information on the generated numbers of~$\Upsilon$. Note that we have used
the moment plug-in estimate for the optimal value of~$\alpha$.

As expected, the variance values of our new estimator and its modifications
are higher than those of Horvitz-Thompson's estimator based on the
non-randomized data. The precision of our basic proposal is practically
acceptable, because, according to the simulations, the corresponding sample
standard deviation of the estimates has gone up  by a mere 60\,\% in
comparison with the Horvitz-Thompson estimate for $M=60\,000$; this results
is  quite reasonable, taking into account that~$Y$ is a very sensitive
variable. Notice, however, that the modification using the knowledge of the
values of~$\Upsilon_{i}$ leads to a substantial reduction in variance.
Thus, while mildly relaxing respondents' privacy on the one hand but still
keeping secret the true response because the true value of the sensitive
variable is never reported, this modification provides estimates whose
precision is comparable with directly surveying under zero non-response. On
the other hand, the high variability of the estimates, even the presence of
negative estimates for the mean wages, shows that the modification using
the switching questions described in sec.~\ref{switchingQuestions} is only
a theoretical exercise and cannot be recommended for practical use. Its
improvement remains an open question.

Comparing contents of all tables, we can see that the mean has practically
not changed; however, the  expected decrease occurs in the variability of
the estimates, of about 9\%, which shows that it pays ``to tune up'' the
procedure and its parameters according to the given problem and potential
data.

Both results of sec.~\ref{application2SRS} and  simulations show that
variance of estimators can be greatly reduced by choice of bounds~$m$
and~$M$. We see that for low value of the upper bound $M=40,000$ the
proposed estimators are competitive even with Horvitz-Thompson estimator.
It follows from the bias formula~\ref{vychyleni} that approximately
unbiased  estimators with low variance can be constructed if we use prior
information on population quantiles for choice of bounds~$m$ and~$M$.
Optimal choice of bounds with respect to the minimization of the mean
square error is field of further research.

\begin{table}[H]
\begin{center}
  \setlength{\extrarowheight}{5pt}
  \begin{tabular}{|l|r|r|r|r|r|}
  \cline{3-6}
     \multicolumn{2}{l} {}          &
     \multicolumn{2}{|c|}{$N=200$}  &
     \multicolumn{2}{|c|}{$N=400$}  \\
  \hline
     \multicolumn{1}{|c}{Estimator} &
     \multicolumn{1}{l} {}          &
     \multicolumn{1}{|c|}{$n=20$}   &
     \multicolumn{1}{|c|}{$n=50$}   &
     \multicolumn{1}{|c|}{$n=20$}   &
     \multicolumn{1}{|c|}{$n=50$}   \\
  \hline
   $\overline{t}_{HT}$
         &  mean  &  24.270  &  24.272  &  24.287  &  24.288 \\
         &  sd    &   2.782  &   1.757  &   2.773  &   1.758 \\
  \hline
   $\overline{t}^R_{AV,(m,M)}$
         &  mean  &  23.189  &  23.192  &  23.203  &  23.205 \\
         &  sd    &   3.687  &   2.333  &   3.690  &   2.336 \\
  \hline
   $\overline{t}^R_{AV,\alpha}$
         &  mean  &  23.192  &  23.194  &  23.206  &  23.207 \\
         &  sd    &   3.000  &   1.897  &   3.001  &   1.902 \\
  \hline
   $\overline{t}^R_{AV,\alpha_{opt}}$
         &  mean  &  23.192  &  23.194  &  23.206  &  23.207 \\
         &  sd    &   2.965  &   1.875  &   2.966  &   1.880 \\
  \hline
   $\overline{t}^R_{AV,T}$
         &  mean  &  23.185  &  23.189  &  23.199  &  23.202 \\
         &  sd    &   6.066  &   3.836  &   6.068  &   3.837 \\
  \hline
  \end{tabular}
  \medskip
  \caption{Numerical results of simulations. The mean estimated salaries
           (in $10^3$~CZK) and the corresponding sample standard deviations
           (in $10^3$~CZK)  for different population sizes~$N$ and sample
           sizes~$n$. Random numbers~$\Upsilon_i$ are generated from the
           uniform distribution on the interval
           $[m, M]=[7,000; 40,000]$,
           $T=30,000$,
           $\alpha=0.75$,
           $\alpha_{opt}=0.72$,
           $1,000$ simulated populations,
           $1,000$ replications of each.}
          \label{tabRes1}
\end{center}
\end{table}

\begin{table}[H]
\begin{center}
  \setlength{\extrarowheight}{5pt}
  \begin{tabular}{|l|r|r|r|r|r|}
  \cline{3-6}
     \multicolumn{2}{l} {}          &
     \multicolumn{2}{|c|}{$N=200$}  &
     \multicolumn{2}{|c|}{$N=400$}  \\
  \hline
     \multicolumn{1}{|c}{Estimator} &
     \multicolumn{1}{l} {}          &
     \multicolumn{1}{|c|}{$n=20$}   &
     \multicolumn{1}{|c|}{$n=50$}   &
     \multicolumn{1}{|c|}{$n=20$}   &
     \multicolumn{1}{|c|}{$n=50$}   \\
  \hline
   $\overline{t}_{HT}$
         &  mean  &  24.297  &  24.301  &  24.288  &  24.290 \\
         &  sd    &   2.773  &   1.758  &   2.813  &   1.779 \\
  \hline
   $\overline{t}^R_{AV,(m,M)}$
         &  mean  &  23.983  &  23.984  &  23.965  &  23.974 \\
         &  sd    &   5.530  &   3.501  &   5.529  &   3.495 \\
  \hline
   $\overline{t}^R_{AV,\alpha}$
         &  mean  &  23.974  &  23.976  &  23.956  &  23.965 \\
         &  sd    &   4.401  &   2.786  &   4.398  &   2.780 \\
  \hline
   $\overline{t}^R_{AV,\alpha_{opt}}$
         &  mean  &  23.976  &  23.977  &  23.958  &  23.967 \\
         &  sd    &   4.164  &   2.637  &   4.161  &   2.631 \\
  \hline
   $\overline{t}^R_{AV,T}$
         &  mean  &  23.991  &  23.992  &  23.973  &  23.982 \\
         &  sd    &   9.066  &   5.729  &   9.067  &   5.726 \\
  \hline
  \end{tabular}
  \medskip
  \caption{Numerical results of simulations. The mean estimated salaries
           (in $10^3$~CZK) and the corresponding standard deviations (in
           $10^3$~CZK)  for different population sizes~$N$ and sample
           sizes~$n$. Random numbers~$\Upsilon_i$ are generated from the
           uniform distribution on the interval
           $[m, M]=[7,000; 60,000]$,
           $T=45,000$,
           $\alpha=0.75$,
           $\alpha_{opt}=0.59$,
           $1,000$ simulated populations,
           $1,000$ replications of each.}
          \label{tabRes2}
\end{center}
\end{table}

\begin{table}[H]
\begin{center}
  \setlength{\extrarowheight}{5pt}
  \begin{tabular}{|l|r|r|r|r|r|}
  \cline{3-6}
     \multicolumn{2}{l} {}          &
     \multicolumn{2}{|c|}{$N=200$}  &
     \multicolumn{2}{|c|}{$N=400$}  \\
  \hline
     \multicolumn{1}{|c}{Estimator} &
     \multicolumn{1}{l} {}          &
     \multicolumn{1}{|c|}{$n=20$}   &
     \multicolumn{1}{|c|}{$n=50$}   &
     \multicolumn{1}{|c|}{$n=20$}   &
     \multicolumn{1}{|c|}{$n=50$}   \\
  \hline
   $\overline{t}_{HT}$
         &  mean  &  24.275  &  24.273  &  24.299  &  24.299\\
         &  sd    &   2.765  &   1.739  &   2.753  &   1.737\\
  \hline
   $\overline{t}^R_{AV,(m,M)}$
         &  mean  &  24.138  &  24.140  &  24.158  &  24.168\\
         &  sd    &   6.911  &   4.372  &   6.921  &   4.378\\
  \hline
   $\overline{t}^R_{AV,\alpha}$
         &  mean  &  24.145  &  24.146  &  24.165  &  24.174\\
         &  sd    &   5.962  &   3.770  &   5.950  &   3.767\\
  \hline
   $\overline{t}^R_{AV,\alpha_{opt}}$
         &  mean  &  24.143  &  24.145  &  24.163  &  24.173\\
         &  sd    &   5.404  &   3.417  &   5.398  &   3.417\\
  \hline
   $\overline{t}^R_{AV,T}$
         &  mean  &  24.136  &  24.137  &  24.156  &  24.165\\
         &  sd    &  13.018  &   8.236  &  13.036  &   8.244\\
  \hline
  \end{tabular}
  \medskip
  \caption{Numerical results of simulations. The mean estimated salaries
           (in $10^3$~CZK) and the corresponding standard deviations (in
           $10^3$~CZK)  for different population sizes~$N$ and sample
           sizes~$n$. Random numbers~$\Upsilon_i$ are generated from the
           uniform distribution on the interval
           $[m, M]=[7,000; 80,000]$,
           $T=45,000$,
           $\alpha=0.75$,
           $\alpha_{opt}=0.53$,
           $1,000$ simulated populations,
           $1,000$ replications of each.}
          \label{tabRes3}
\end{center}
\end{table}

\vfill\eject

\vfill\eject

\begin{figure}[H]
\begin{center}
  \includegraphics[scale=1]{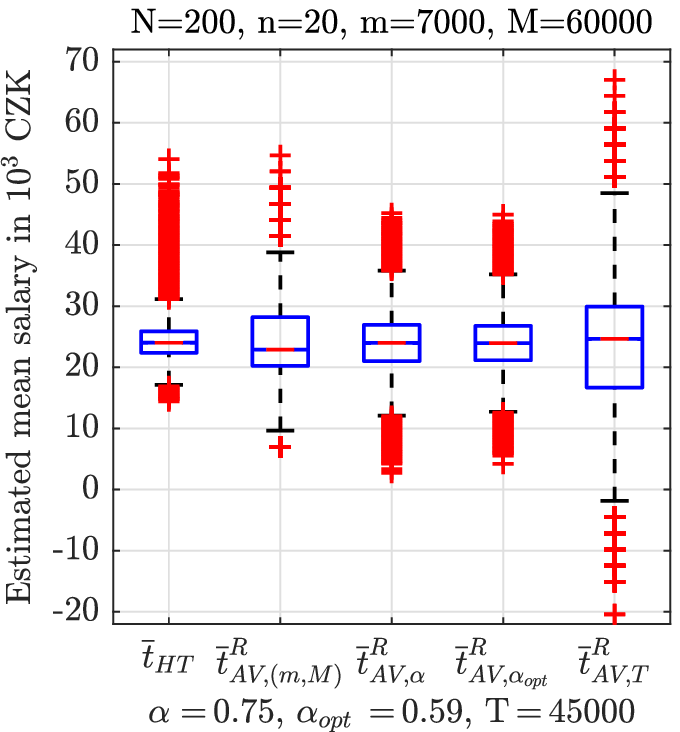}
  \includegraphics[scale=1]{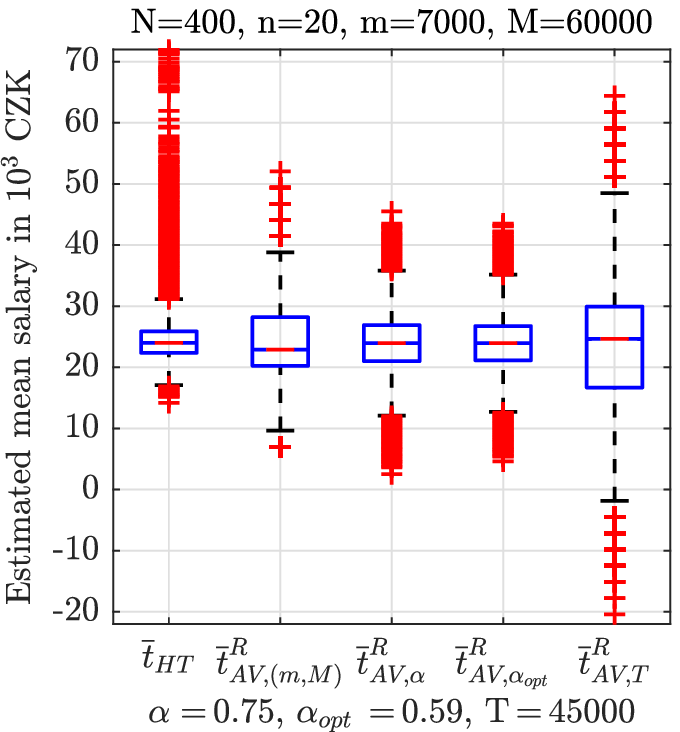}

  \bigskip
  \includegraphics[scale=1]{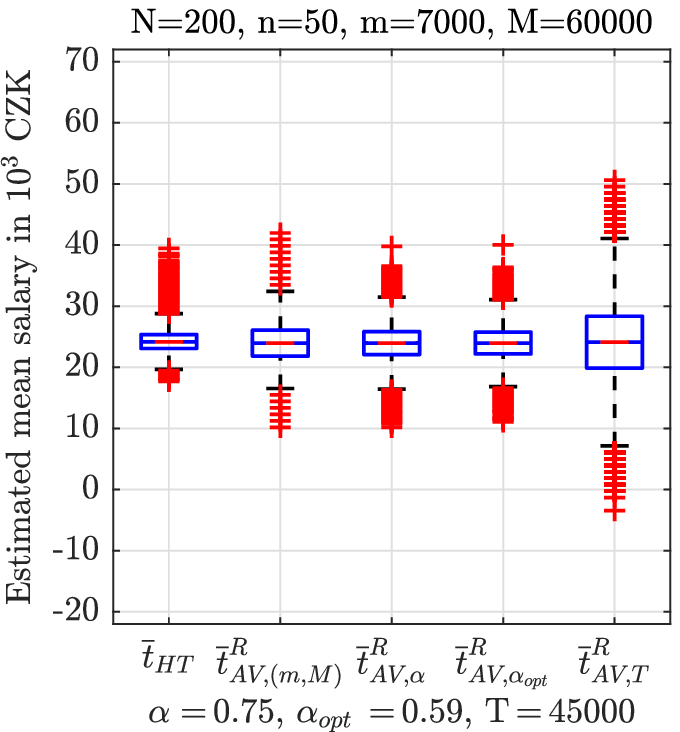}
  \includegraphics[scale=1]{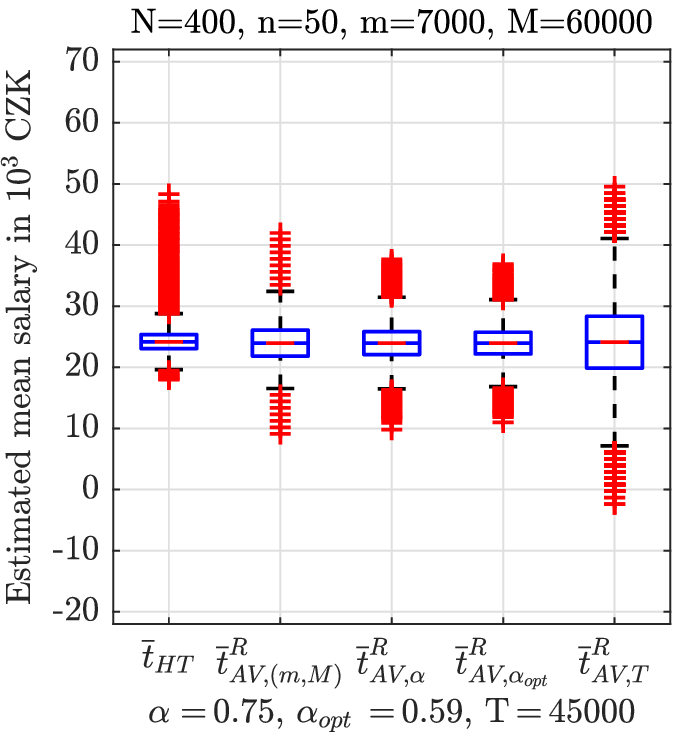}
  \caption{Behavior  of considered estimators applied to different sizes of
           the population~$N$ and sample sizes~$n$; $(m,M)=(7,000; 6,000)$,
           $T=45,000$, $\alpha=0.75$ and $\alpha_{opt}=0.59$.}
  \label{results60}
\end{center}
\end{figure}

\vfill\eject

\section{Conclusions}\label{conclusions}

The purpose of this paper is to present a new randomized response technique
possessing two attractive properties, namely:

\begin{itemize}
  \item It is simple to use.
  \item It provides a high level of anonymity to the respondent.
\end{itemize}

\noindent
Though a quantitative estimate is the final end, the respondent is only
asked for a qualitative response. Two modifications are discussed as well.
The suggested estimators are based on the values of pseudorandom numbers
generated by the respondents, which are used for masking sensitive
information.

A small disadvantage of the suggested method may, for some respondents, be
a feeling of infringement on their privacy  due to an extrinsic
device/technique being used for generating the random numbers. This problem
is of mainly a  psychological nature and can,  at least partially, be
resolved by a proper explanation of the approach by the interviewer.
Unfortunately, all currently used RRT procedures suffer, to a certain
extent, from the same problem -- see, for example, the thorough discussion
in Chaudhuri (2017) and Chaudhuri and Christofides (2013).

The first modification assumes that not only the respondent, but also the
interviewer knows the generated random number that masks the true value of
the response. The second modification makes use of switching questions with
the aim to make the survey less embarrassing for respondents in certain
specific situations. For all suggested RRT procedures, we show their
unbiasedness, and derive the corresponding variance for the
Horvitz-Thompson's type estimator under the simple random sampling without
replacement. The optimal values of the tuning parameters enabling us to
minimize the variance of the suggested procedures are also discussed. The
first modification seems to be especially promising because we have shown
that knowing the random number and properly setting the tuning parameters
can sufficiently increase the precision of the estimator. For the second
modification, it is good to know that it would not work in
practice. On the other hand, we admit that, for some readers, the suggested
modifications may be of interest, even if only from the theoretical point
of view.

As a technical tool, two auxiliary measures are proposed, called the mean
relative concentration measure of the values of~$Y$ around the center of
interval $(0,M)$, and the proximity measure of the population mean to the
center of interval $(0,M)$. With the aid  of these measures we can explain
why, and especially how, the suggested RRTs increase the variance of the
estimators of~$t_Y$ and~$\overline{t}_Y$ for symmetrical distributions;
distributions closely concentrated around their centers; or uniform
distributions.

We would like to summarize the merits of the method proposed in this paper.
In our opinion, we are bringing progress in this field. The first advantage
is that our method is easy to implement because there exist many more or
less easily available online/offline generators of random numbers from the
uniform distribution. If the main goal of a survey is to estimate a
continuous random variable with a large span, like income or personal
wealth then, for the ``classical RRT methods'' described in
sec.~\ref{prehled}, we need to design a very large deck of cards to mask
the true values of the surveyed variable. For example, if we assume an
income range from 7,000~CZK to 60,000~CZK, as is reasonable in our example,
the number of cards needed  for Eriksson's RRTs, provided the income values
are rounded to 1,000~CZK, is 54. If the rounding step is 500~CZK, then 107
cards are needed. Finally, if the rounding step is 100 CZK, then 531 cards
are needed. Manipulations with such a large deck of cards can be cumbersome
for both the respondent and the interviewer. Even if the span of the
surveyed variable is not very large, it is not easy to find precise
instructions, or algorithms, concerning how to design the corresponding
deck of cards. Difficulties with this design may pose a problem for the
field survey statisticians, discouraging them from the use of such RRTs.

Our technique also shares the ease of use with Eriksson's technique. Unlike
within Chaudhuri's approach, which requires quite demanding arithmetic
operations from the respondent, each respondent only states whether his/her
true income is higher than a certain number. Let us point out that the
respondent never reports the true value of the variable. In our original
proposal, described in sec.~\ref{newProcedure}, the interviewer moreover
does not  know the value of $\Upsilon$. Thus, the privacy of respondent
is protected better than in Eriksson's approach,  which intrudes on the
privacy of the respondents to a certain extent. Indeed, if the value
reported by a respondent differs from any of $x_{t}$, the interviewer
learns about the true value of the sensitive variable.

Finally, note that we find rather problematic any comparison of our
approach with the methods employed by Eriksson or Chaudhuri, because their
performance strongly depends on the choice of the cards used. In our
opinion, it is tricky to design a deck of cards for a continuous variable
with a high range, such as the income in the Czech Republic, and a reliable
estimator of this type with an acceptably small variance value would need
an excessively large deck of cards.

\bigskip\noindent
\textbf{Acknowledgements:} The work was supported by grant GA\v{C}R
P403/19/02773S.

\renewcommand{\refname}{References}

\end{document}